\documentclass[12pt]{article}
\usepackage{geometry}                
\usepackage{graphicx}
\usepackage{amssymb}
\usepackage{amsmath}
\usepackage{epstopdf}
\let\a=\alpha \let\b=\beta \let\g=\gamma

\let\w=\omega

\def\ba{\begin{array}}
\def\ea{\end{array}}

\def\td{\tilde}

\def\dalemb#1#2{{\vbox{\hrule height .#2pt
        \hbox{\vrule width.#2pt height#1pt \kern#1pt
                \vrule width.#2pt}
        \hrule height.#2pt}}}

\def\R{{{\cal R}}}
\def\F{{{\cal F}}}

\def\Lag{{\mathcal{L}}}

\def\Real{{\rm Re}\,}
\def\Imag{{\rm Im}\,}

 \newcommand{\be}{\begin{equation}}
\newcommand{\ee}{\end{equation}}
 \newcommand{\bal}{\begin{align}}
  \newcommand{\eal}{\end{align}}
 \newcommand{\ben}{\begin{equation*}}
\newcommand{\een}{\end{equation*}}
\newcommand{\bea}{\begin{eqnarray}}
\newcommand{\eea}{\end{eqnarray}}
\newcommand{\bean}{\begin{eqnarray*}}
\newcommand{\eean}{\end{eqnarray*}}
\newcommand{\bes}{\begin{subequations}}
\newcommand{\ees}{\end{subequations}}

\begin{document}

\begin{titlepage}
\bigskip
\rightline{}

\bigskip\bigskip\bigskip\bigskip
\centerline {\Large \bf { Zero Temperature Limit of  }}
\bigskip
\centerline{\Large \bf Holographic Superconductors}
\bigskip\bigskip
\bigskip\bigskip

\centerline{\large  Gary T. Horowitz and Matthew M. Roberts}
\bigskip\bigskip
\centerline{\em Department of Physics, UCSB, Santa Barbara, CA 93106}
\centerline{\em  gary@physics.ucsb.edu, matt@physics.ucsb.edu}
\bigskip\bigskip
\begin{abstract}
We consider holographic superconductors whose bulk description consists of gravity minimally coupled to a Maxwell field and  charged scalar field with general potential. We give an analytic argument that there is no ``hard gap": the real part of the conductivity at low frequency remains nonzero (although typically exponentially small) even at zero temperature. We also numerically construct the gravitational dual of the ground state of some holographic superconductors. Depending on the charge and dimension of the condensate, the infrared theory can have emergent conformal or just Poincare symmetry. In all cases studied, the area of the horizon of the dual black hole goes to zero in the extremal limit, consistent with a nondegenerate ground state.

\end{abstract}
\end{titlepage}

\tableofcontents
\baselineskip 16pt

\setcounter{equation}{0}
\section{Introduction}

Over the past couple years, it has been shown that various familiar properties of condensed matter systems can be reproduced using a theory of gravity with anti de Sitter boundary conditions \cite{Hartnoll:2009sz}. This is a new application of the AdS/CFT correspondence (which  is  being called AdS/CMT). There are, by now, several bulk theories which describe superconductivity on the boundary \cite{Gubser:2008px,Hartnoll:2008vx,Gubser:2008wv,Denef:2009tp,Gubser:2009qm,Gauntlett:2009dn}
. They all contain charged black hole solutions which develop nontrivial hair at low temperatures. The simplest theory (and first one discussed) contains gravity minimally coupled to a Maxwell field and charged scalar with potential $V(\psi)$. Although various properties of this theory have been discussed \cite{Hartnoll:2009sz,Herzog:2009xv}, the actual zero temperature ground state has remained mysterious. In this paper we clear up this mystery by constructing the extremal limit of the hairy black holes which are dual to the superconductor. 

 One of the main open questions  concerns the behavior of the conductivity at low frequency. It is known that at at low (but  nonzero) temperature there is a pronounced gap in $\Real \sigma(\w)$ which numerically appears to be $\Real \sigma(\w) \sim e^{-\Delta/T}$ for some constant $\Delta$. The open question is whether the conductivity is strictly zero in the limit $T=0$. We will present a simple analytic argument that the answer is no, for all $V(\psi)$. This follows from the fact that the conductivity can be  related to a reflection coefficient $\R$ in  a simple one dimensional scattering problem off a positive potential, where  the incident energy is $\w^2$.  $\Real \sigma(\w) =0 $ requires $|\R|=1$. We will see that in all cases the potential vanishes at the horizon, so  there is always a small probability for transmission. This shows $\sigma(\w)$ will be  small but nonzero at low frequency and zero temperature. As one lowers the temperature, the Schr\"odinger potential becomes higher and wider reproducing the earlier result $\Real \sigma(\w) \sim e^{-\Delta/T}$ for a range of $T>0$. However the potential saturates at $T=0$ causing $\Real \sigma(\w) >0$ even at $T=0$.  In some cases with emergent conformal symmetry in the infrared, the Schr\"odinger potential never gets very high or wide and $\Real \sigma(\w)$ does not show a strong gap, in agreement with \cite{Gubser:2008wz,Gubser:2009gp}. This Schr\"odinger picture of the conductivity will also allow us to clarify the spikes that were found in $\Real \sigma(\w)$ in \cite{Horowitz:2008bn}.
 
In addition, we numerically construct the $T=0$ limit of the hairy black holes for the case $V = m^2 |\psi|^2$.
One's first thought is that the extremal limit should resemble extremal Reissner-Nordstrom AdS with a degenerate horizon at nonzero radius. However, this would be highly problematic from the standpoint of the dual superconductor, since a nonzero horizon area implies a highly degenerate ground state. Indeed,  it has recently been shown that this model cannot have a smooth degenerate horizon at nonzero radius \cite{FernandezGracia:2009em}. 

When  $V = m^2 |\psi|^2$, the bulk theory depends on two parameters, the mass $m$ and charge $q$  of the complex scalar field. We consider two cases in detail: $m=0$; and $m^2 \le 0$ and $q^2 > |m^2|/6$. In both cases, the radius of the horizon goes to zero in the extremal limit, consistent with the idea of a unique ground state. In the first case, the solution approaches $AdS_4$ near the horizon implying that conformal invariance is restored in the infrared. In the second case, Poincare invariance (but not conformal invariance) is restored in the infrared\footnote{Some early indications of emergent Poincare symmetry were found in \cite{Gubser:2008pf}}. The solutions in the latter case have  a null curvature singularity at the extremal horizon.  Since they arise as the extremal limit of black holes with smooth horizons, these singular solutions are still physical.  (They are perhaps analogous to Dp-brane metrics for $p<6$.) When $m^2 =0$, the solutions are nonsingular and describe static charged scalar solitons. However, even though the 
curvature remains finite at the extremal horizon, in most cases derivatives of the curvature will diverge. For one special value of $q$, the solution is completely smooth across the horizon.

The fact that the extremal solution appears to behave qualitatively differently for very small $q$ is consistent with the fact that there are two different instabilities which produce the scalar hair \cite{Hartnoll:2008kx}.  We will  present analytic solutions for the leading small $r$ behavior for three more cases: (1) $m^2 < 0 $, $q^2$ small; (2) $q=0$; and (3) $m^2 > 0 $. We have not been able to show that these near horizon solutions match onto the standard asymptotic boundary conditions. 
When $q=0$, the scalar hair cannot carry charge, and all the charge remains on the extremal black hole. In all other cases, the extremal black hole appears to have zero charge. This is expected since a nonzero charge on a black hole with zero radius would produce a diverging electric field. If the scalar field carries any charge, there will be a superradiant instability which causes the black hole to lose its charge. (Quantum mechanically, the diverging electric field pair creates charged quanta.)

As this paper was being completed, \cite{Gubser:2009cg} appeared which also discusses the ground states of holographic superconductors. However, that paper focusses on the zero temperature condensates for W-shaped $V(\psi)$ where $V$ has additional extrema. The only overlap with our discussion is in section 5.3 where we consider the possibility of  Lifshitz spacetimes emerging in the infrared.

\setcounter{equation}{0}
\section{Preliminaries}
In this section we review the bulk description of the simplest holographic superconductor. For more details, see \cite{Hartnoll:2009sz,Herzog:2009xv}. We also review the condition for scalar hair to arise at low temperature \cite{Denef:2009tp}.

\subsection{Holographic superconductors}

We begin with the following four dimensional action describing gravity minimally coupled to a Maxwell field and charged scalar:
\be\label{eq:bulktheory}
\Lag = R + \frac{6}{L^2} - \frac{1}{4} F^{\mu\nu} F_{\mu\nu} 
- |\nabla \psi - i q A \psi |^2 -V( |\psi|) \,.
\ee
As usual we are writing $F=dA$, the cosmological constant is $-3/L^2$, and $m,q$ are the mass and charge of the scalar field. We are interested in plane symmetric solutions, so we set
\be\label{metric}
 ds^2=-g(r) e^{-\chi(r)} dt^2+{dr^2\over g(r)}+r^2(dx^2+dy^2)
\ee
\be
A=\phi(r)~dt, \quad \psi = \psi(r)
\ee
We can choose a gauge in which $\psi$ is real and work in units with $L=1$. The equations of motion are:
\be \psi''+\left(\frac{g'}{g}-\frac{\chi'}{2}+\frac{2}{r} \right)\psi' +\frac{q^2\phi^2e^\chi}{g^2}  \psi  -{V'(\psi)\over 2g}=0\label{psieom}\ee

\be\label{phieom}
\phi''+\left(\frac{\chi'}{2}+\frac{2}{r}  \right)\phi'-\frac{2q^2\psi^2}{g}\phi=0
\ee

\be
\chi'+r\psi'^2+\frac{rq^2\phi^2\psi^2e^\chi}{g^2}=0\label{chieom}
\ee

\be\label{geom}
g' + \left(\frac{1}{r}  - { \chi'\over 2}\right) g+\frac{r\phi'^2e^\chi}{4}- 3r+\frac{rV(\psi)}{2}=0
\ee
These equations are invariant under a scaling symmetry:
\be\label{rescale}
r \to a r \,, \quad (t,x,y) \to (t,x,y)/a \,, \quad g \to a^2 g \,, \quad \phi \to a \phi
\,.
\ee
When the horizon is at nonzero $r$, this can be used to set $r_+ = 1$.  These equations are also invariant under 
\be\label{rescalet}
e^\chi \to a^2 e^\chi, \quad  t\to at, \quad \phi \to \phi/a
\ee
This symmetry can be used to set $\chi =0$ at the boundary at infinity, so the metric takes the standard AdS form asymptotically.
At large radius
\be 
\phi = \mu -{\rho\over r}, \qquad \psi ={\psi^{(\lambda)}\over r^\lambda}+{\psi^{(3-\lambda)}\over r^{3-\lambda}}.
\ee
where $\lambda = (3 +\sqrt{9+4m^2})/2 $. In the boundary CFT, $\mu$ is the chemical potential, $\rho$ is the charge density, and $\lambda$ is the scaling dimension of the operator dual to $\psi$. We want this operator to condense without being sourced, so we are only interested in solutions where $\psi$ is normalizable. This typically requires setting $\psi^{(3-\lambda)} = 0$.

\subsection{Condition for instability}

To see when one expects hairy black holes at low temperature, one can study
 linearized perturbations of the extremal Reisner-Nordstrom AdS (RN-AdS) black hole. Using the scaling symmetry (\ref{rescale}) to set the horizon radius to one, the general RN-AdS  solution is given by
\be \chi=\psi=0,\quad g=r^2-\frac{1}{r}\left(1+\rho^2/4\right)+\rho^2/4r^2, \quad \phi=\rho\left(1-1/r\right)\label{RN}
\ee 
The temperature of the black hole (\ref{metric}) is 
\be\label{temp}
 T=\frac{\left[g' (g~e^{-\chi})' \right]^{1/2}}{4\pi}|_{r=r_+}
\ee
For AdS-RN, this is $T=(12-\rho^2)/16\pi$, so the extremal limit is $\rho=2\sqrt{3}$. The near-horizon limit of this solution is $AdS_2\times \mathbb{R}^2$,

\be ds^2=-6(r-1)^2 dt^2+\frac{dr^2}{6(r-1)^2}+dx^2+dy^2, \quad \phi=2\sqrt{3}(r-1)
\ee

Plugging this into the scalar wave equation (\ref{psieom}), dropping the $2/r$ as it is negligible compared to the divergence of $g'/g$, and changing variables $\td r=r-1$, we  recover a wave equation for  $AdS_2$ with a new effective mass,
\be\psi_{,\td r \td r}+\frac{2}{\td r}\psi_{,\td r}-\frac{m^2_{eff}}{\td r^2}\psi=0, \qquad ~m^2_{eff}=\frac{m^2-2q^2}{6}
\ee
The instability to form scalar hair at low temperature then is just the instability of scalar fields below the Breitenlohner-Freedman (BF) bound for $AdS_2$: $m^2_{BF} = -1/4$. Thus the condition for instability is 
\be\label{unstable}
 m^2-2q^2<-3/2\ee
Of course, the mass must be above the four-dimensional BF bound,  $m^2 > -9/4$.

\setcounter{equation}{0}
\section{Conductivity}

In this section we reformulate the calculation of the conductivity in a holographic superconductor in terms of a one dimensional Schr\"odinger problem. The conductivity will be simply related to a reflection coefficient\footnote{The reformulation of key equations in terms of an equivalent Schr\"odinger problem has been a useful tool in many applications of AdS/CFT (see, e.g., \cite{Faulkner:2009wj}) but to our knowledge, it has not yet been directly applied to the conductivity. Our relation between the conductivity and the reflection coefficient is perhaps analogous to the relation between the shear viscosity and the black hole absorption coefficient \cite{Policastro:2001yc}.}. This approach provides a simple intuitive understanding of all the qualitative features of the conductivity that were seen at nonzero temperature, and allows us to extend them to $T=0$. In particular, 
it was shown in \cite{Hartnoll:2008kx} that there is a pronounced gap in the conductivity at low frequency and low temperature. The real part of the conductivity is exponentially suppressed, and appeared to satisfy Re $\sigma \sim e^{-\Delta/T}$. These results were obtained numerically, and suggested that the conductivity should strictly vanish at zero temperature. We will see that this is not the case. These holographic superconductors do not have a ``hard gap". This result is consistent with the fact that the specific heat obeys a power law and is not exponentially suppressed at low temperature\footnote{This was first noticed in \cite{Hartnoll:2008kx} and later studied in a different bulk model for a holographic superconductor \cite{Peeters:2009sr}.}. 

\subsection{Conductivity in terms of a reflection coefficient}

To obtain the conductivity, we must solve for a linearized perturbation of the vector potential. Assuming translational symmetry and harmonic time dependence, the perturbation satisfies \cite{Hartnoll:2008kx}
\be
A_x'' + \left[\frac{g'}{g} - \frac{\chi'}{2} \right] A_x'
+ \left[\left(\frac{\w^2}{g^2} - \frac{\phi'^2}{g} \right) e^{\chi} - \frac{2 q^2
\psi^2}{g} \right] A_x  =  0 \,. \label{eq:ax}
\ee
This equation can be simplified by introducing a new radial variable 
\be
dz = {e^{\chi/2}\over g} dr
\ee
At large $r$, $dz  = dr/r^2$, and we can choose the additive constant so that $ z = -1/r$.
Since $g$ vanishes at least linearly at a horizon and $\chi$ is monotonically decreasing, the horizon corresponds to $z=-\infty$. In terms of $z$, (\ref{eq:ax}) takes the form of a standard Schr\"odinger equation:
\be\label{schr}
-A_{x,zz} + V(z) A_x = \omega^2 A_x
\ee
with
\be\label{potential} 
V(z) = g[\phi_{,r}^2  + 2q^2 \psi^2  e^{-\chi}]
\ee
From the known asymptotic behavior of the solution near infinity, we can determine the behavior of the potential. 
Near $z=0$, $V(z) = \rho^2 z^2 + 2(q\psi^{(\lambda)})^2 z^{2(\lambda -1)}$. So the potential vanishes if the dimension of the condensate, $\lambda$, is greater than one, $V(0)$ is a nonzero constant if $\lambda =1$, and $V(z)$ diverges if $1/2 < \lambda < 1$. We will show below that the potential always vanishes at the horizon.

We want to solve (\ref{schr})  with ingoing wave boundary conditions at $z = -\infty$. The easiest way to do this is to first extend the definition of the potential to all $z$ by setting $V=0$ for $z>0$. Now an incoming wave from the right  will be partly transmitted and partly reflected by the potential barrier. Since the transmitted wave is purely ingoing at the horizon, this satisfies our desired boundary conditions.  Writing the solution for $z>0$ as $A_x = e^{-i\omega z} + \R e^{i\omega z}$, we clearly have $A_x(0) = 1+\R$ and $A_{x,z}(0) = - i\omega(1-\R)$. As shown in \cite{Hartnoll:2008kx}, if $ A_x = A_x^{(0)} + A_x^{(1)}/r$, then the conductivity is
\be\label{oldcond}
\sigma(\omega) = -{i\over \w} {A_x^{(1)}\over A_x^{(0)}}
\ee
In terms of $z$, $ A_x^{(1)} = - A_{x,z} (0)$, so
\be\label{cond}
\sigma(\omega) =  {1-\R\over 1+\R}
\ee
The  conductivity is directly related to the reflection coefficient, with the frequency simply giving the incident energy! The qualitative behavior of $\sigma(\omega)$ is now clear. Let us first assume that $\lambda\ge 1$ so that $V$ is bounded. At frequencies below the height of the barrier, the probability of transmission will be  small, $\R$ will be  close to one, and $\sigma(\omega)$ will be  small. At frequencies above the height of the barrier, $\R$ will be very small and $\sigma(\omega) \sim 1$ (the normal state value). Clearly the size of the gap in $\sigma(\omega)$ is set by the height of the barrier: $\w_g \sim \sqrt {V_{max}}$. 
The case $1/2< \lambda < 1$ is qualitatively similar. Even though the potential is not bounded, $\sqrt{V}$ is integrable, so there is still tunneling through the barrier. 

The key point is that $\Real \sigma(\omega)$ is never strictly zero. That would require a potential which remains nonzero at the horizon. To illustrate this, consider  a $\lambda=1$ scalar in the probe limit, where the (zero temperature) metric is just AdS in Poincare coordinates. In this case, the potential (\ref{potential}) does not have the $\phi'^2$ term, and $\chi=0$. Assuming $\psi = \psi^{(1)}/r$ everywhere (which is close to the numerical result),  $V(z) = V_0$ for $z<0$ with $V_0 = 2[q \psi^{(1)}]^2$. One can easily compute the reflection coefficient for this potential and find
\be
\sigma(\w) = {i\over\w} \sqrt{V_0 - \w^2} \qquad {\rm for}\  \w^2 < V_0
\ee
In other words, $\sigma(\w)$ is purely imaginary and the real part is strictly zero. (The imaginary part has a simple pole corresponding to the infinite DC conductivity.) This reproduces the analytic form discussed in \cite{Hartnoll:2008vx}.  However, since $\psi$ is diverging near the horizon, the probe approximation breaks down. We will see below that when backreaction is included, the Schr\"odinger potential will vanish there.

This entire discussion applies to  black holes with nonzero temperature as well as  extremal solutions. For a nonextremal solution, all fields are finite at the horizon, and $g(r_+) = 0$, so the horizon again corresponds to $z = -\infty$ and the potential clearly vanishes there. As one lowers the temperature the potential becomes both higher and wider so the exponential suppression increases  (see Fig.  1).  This is why the earlier papers had seen the approximate behavior Re $\sigma\sim e^{-\Delta/T}$. However, as $T\rightarrow 0$, the potential approaches a finite, limiting form which still vanishes at $z = -\infty$. This means that there will always be nonzero tunneling probability and hence a nonzero conductivity.

\begin{figure}
\begin{center}\label{q10potential}
\includegraphics[width=.7\textwidth]{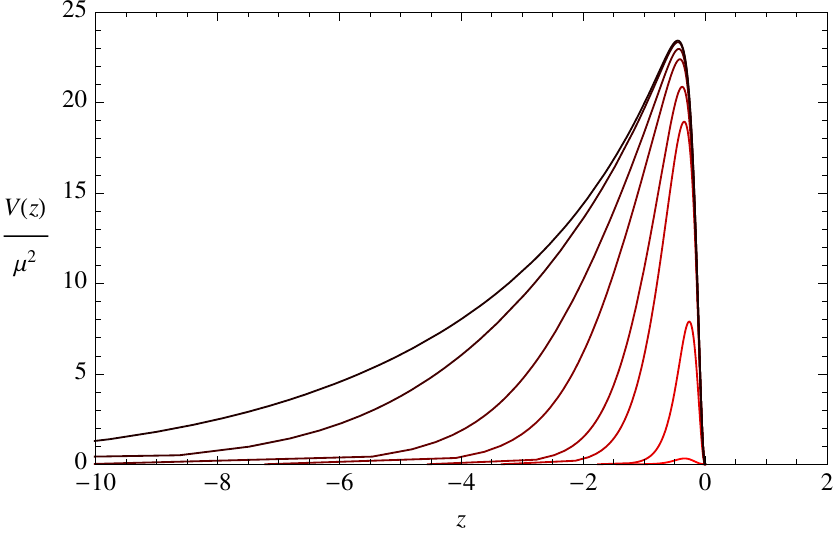}\caption{Schr\"odinger potential for $\lambda=2, q=10$. The potential increases as $T/T_c$  is lowered from one to zero. See section 4.2 for more details on this solution.}\label{pot}
\end{center}\end{figure}

It has recently been shown that for  a W-shaped potential, i.e., either general quartic potentials \cite{Gubser:2008wz}
 or specific theories obtained from string theory truncations \cite{Gubser:2009gp},  the conductivity is not suppressed at low frequency and zero temperature. Instead, $\Real \sigma(\w)$ is a power law in $\w$ with a coefficient of order one. This is a result of the fact that in these models, the Schr\"odinger potential never gets very high. The solution usually approaches $AdS$ near the horizon with a (possibly) shifted cosmological constant.  In this case,  $\chi$ and $\phi$ are constant near the horizon, and typically of order one. As an example, we give the zero temperature Schr\"odinger potential for the case $V(\psi)=0$ and $q=1$ in Figure 2.  As we discuss in the next section, this solution approaches AdS near the horizon. We will show below that $\Real \sigma(\w)$ is a power law in $\w$ even in some cases which do not approach AdS near the horizon. In these cases, the coefficient in front of the power law can be exponentially small.
  
\begin{figure}
\begin{center}\label{masslesspotential}
\includegraphics[width=.7\textwidth]{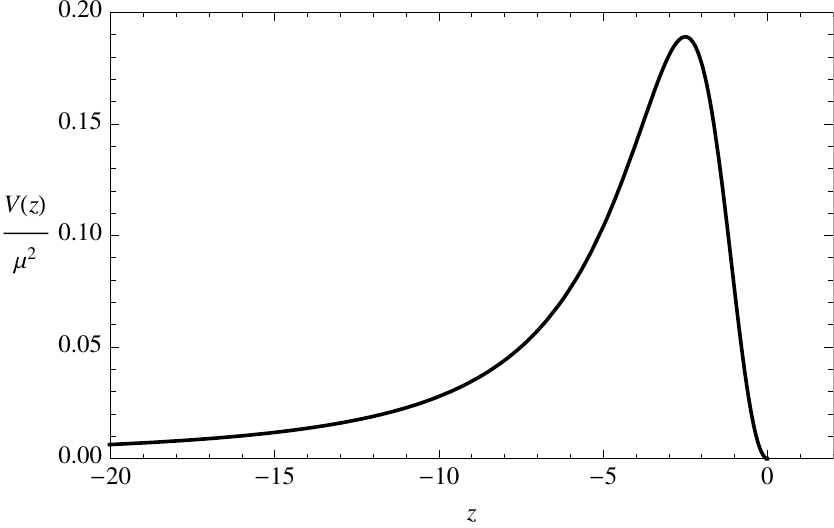}\caption{Schr\"odinger potential for $\lambda=3, q=1$ at $T=0$. Note that the potential does not rise as high as it does in figure 1. See section 4.1 for more details on this solution.}\label{pot}
\end{center}\end{figure}

From the Kramers-Kronig relations, in order for $\Real \sigma$ to have a delta function at $\omega = 0$ representing the infinite DC conductivity, one needs  ${\rm Im}\, \sigma$ to have a pole at $\omega =0$. It is easy to see that this is indeed the case, for any positive potential $V(z)$ that vanishes at $z=-\infty$.  Imagine solving (\ref{schr}) with $\omega = 0$, and $A_x = 1$ at $z=-\infty$. (This represents the normalizable solution.) Since $A_{x,zz} > 0$, the solution will be monotonically increasing. At $z=0$, $A_x^{(0)} = A_x(0)$ and $A_x^{(1)} = - A_{x,z}(0)$. These are both real and nonzero. From (\ref{oldcond}) it then follows that ${\rm Im}\, \sigma$ has a pole at $\omega =0$.

This approach also explains the spikes in the conductivity that were found in \cite{Horowitz:2008bn}.  At low frequency, the incoming wave from the right is almost entirely reflected. If the potential is high enough, one can raise the  frequency so that about one wavelength fits between the potential and $z=0$. In this case, the reflected wave can interfer destructively with the incident wave and cause its amplitude at $z=0$ to be exponentially small. This produces a spike in the conductivity as seen in \cite{Horowitz:2008bn}. If one can raise the frequency so that two wavelengths fit between the potential and $z=0$ one gets a second spike, etc. More precisely, using standard WKB matching formula,
spikes will occur when there exists  $\omega$ satisfying
\be
\int_{-z_0}^0 \sqrt{\omega^2 - V(z)} dz + {\pi\over 4} = n\pi
\ee
for some integer $n$, where $V(-z_0) = \omega^2$. The spikes were first seen using a probe approximation with $V = m^2 \psi^2$  and letting the mass saturate the BF bound. It is now clear that the spikes will appear in the full backreacted solutions, and for some $m^2$ slightly above the BF bound.
 When the spikes were first seen, it was speculated that they corresponded to vector normal modes of the hairy black hole. It is now clear that they are not true normal modes even at $T=0$, since $A_x$ does not actually vanish at infinity.  The actual modes all have complex frequency and correspond to familiar quasinormal modes. In other words, there are no bound states in this potential (with boundary condition $A_x=0$ at $z=0$) since the potential vanishes at $z=-\infty$.

\subsection{Proof that the Schr\"odinger potential vanishes at the horizon}

Since $V$ is explicitly proportional to $g$, it is clear that if the zero temperature solution is smooth at the horizon, the potential will vanish there, and the conductivity will not be strictly zero at low frequency. This applies to the recent embeddings of holographic superconductors in string theory \cite{Gubser:2009qm,Gauntlett:2009dn}.  However, in general, the extremal limit is not smooth. (We will show this explicitly in the next section.) We know that $ g$ and $\phi$ must vanish at the extremal horizon since they vanish on the horizon for every $T>0$ solution, but in general, $\phi',  \psi$ and $  \chi$ can diverge.

We now show that even for singular solutions, the Schr\"odinger potential always vanishes at the horizon. We will prove this using mostly equations (\ref{phieom}) and (\ref{chieom}) which do not depend on $V(\psi)$, and the boundary conditions that $ g$ and $\phi$ must vanish at the extremal horizon.
We will consider the two terms in (\ref{potential}) separately.  The argument is slightly different if the extremal horizon is at $r_+ = 0 $ or $r_+ >0$ and we will consider both possibilities.  We will use prime to denote $d/dr$.

\subsubsection{$g{\phi'}^2$ must vanish on the horizon}

Suppose $g{\phi'}^2 = k^2 \ne 0$ on the horizon. Then $\phi'= k/\sqrt{g}$ and hence $\phi^{''}= -kg'/2g^{3/2}$. Thus (\ref{phieom}) becomes
\be\label{newphieq}
-{kg' \over 2g^{3/2}} + \left(\frac{\chi'}{2}+\frac{2}{r}  \right){k\over g^{1/2}} -\frac{2q^2\psi^2}{g}\phi=0
\ee
Since (\ref{chieom}) implies that $\chi'$ must be negative,  the only positive term is $2k/rg^{1/2}$. Suppose $r_+ >0$. We claim that this positive term is always dominated by the first term above. Indeed, if they were equal, then $g'/g = 4/r_+$ which implies $g \propto e^{4r/r_+}$ which does not vanish on the horizon. If $g$ does vanish on the horizon, then $g'/g = (\log g)' \gg 4/r_+$. Since the remaining terms are all negative there is no solution.

Now suppose $r_+ = 0$. The positive term is no longer always dominated by the first term, but we can proceed as follows. Since $\phi$ must vanish on the horizon, $g$ must not only vanish, but $g^{-1/2}$ must be integrable. Let us parameterize $g = r^{2\g} $ with $0<\g<1$. Then (\ref{newphieq}) becomes
\be
{2-\g\over r^{1+\g}} + {\chi'\over 2r^\g} - {2q^2 \psi^2 r^{1-3\g}\over 1-\g} = 0
\ee
This can be rewritten:
\be\label{phiconstraint}
2-\g = -{1\over 2} r\chi' + {2q^2 \psi^2 r^{2-2\g}\over 1-\g} 
\ee
Since the two terms on the right hand side are positive, neither can diverge. If the second term on the right is nonzero at the horizon, then $\psi \propto 1/r^{1-\g}$. But substituting this into (\ref{chieom}) we see that this implies that $r\chi'$ diverges at $r=0$.  This contradicts (\ref{phiconstraint}), so the second term on the right must vanish. Thus, $\chi' = -(4-2\g)/r$ which implies $e^\chi \propto  1/r^{4-2\g}$. But this implies that the last term in (\ref{chieom}) is now
\be
\frac{rq^2\phi^2\psi^2e^\chi}{g^2} \propto {\psi^2\over r^{1 + 4\g}}
\ee
Thus $r\chi'$ again diverges at $r=0$ unless $\psi$ vanishes.  To obtain a contradiction, we turn to eq. (\ref{geom}). The dominant term in  this equation (excluding the $V(\psi)$ term) is
$r{\phi'}^2e^\chi/4 \propto 1/r^3$.  This cannot be cancelled by $rV(\psi)/2$ unless $V$ diverged at $\psi =0$. However,  $V(0)$ must be a finite negative number in order to have an asymptotic AdS region. This completes the argument that $g{\phi'}^2$ must vanish on the  horizon, even if it is singular in the extremal limit.

\subsubsection{$g\psi^2 e^{-\chi}$ must vanish on the horizon}

Since $g$ must vanish at the horizon, this term can be nonzero only if $\psi^2 e^{-\chi}$ diverges there. Suppose $r_+ > 0 $. We will get a contradiction using only $\chi' = -r{\psi'}^2$. Since we have dropped the last term in (\ref{chieom}) and  $\chi$ vanishes at infinity, this is a lower estimate for $\chi$. The actual $\chi$ must be even larger. So assume
\be\label{diverge}
\psi^2 e^{-\int_r^\infty\tilde r \psi'^2 d\tilde r} =f
\ee
where $f\rightarrow \infty$ as $r\rightarrow r_+$. Taking the logarithm of both sides and then a derivative, yields
\be
r{\psi'}^2 + {2\psi'\over \psi} - {f'\over f} =0
\ee
Viewing this as a quadratic equation for $\psi'$, we can solve it to obtain
\be\label{contradiction}
r\psi \psi' = -1 \pm \sqrt{1+ r \psi^2 f'/f}
\ee
Since $f\rightarrow \infty$ as $r\rightarrow r_+$, $f'/f \rightarrow -\infty$. It follows that real solutions require $\psi(r_+)=0$. But the right hand side is negative, so integrating (\ref{contradiction}) from $r_+$ to $r_+ + \epsilon$ yields $\psi^2 <0$. This contradiction shows that our assumption that $\psi^2 e^{-\chi}$ can diverge is incorrect.

Now suppose $r_+ =0$.  It turns out that in this case (\ref{diverge}) can be satisfied, e.g., for  $\psi = A(-\log r)^{1/2}$ near $r=0$. However, this requires $\psi$ and $\chi$ to both diverge slowly near $r=0$.   In particular, $\chi' $ must diverge more slowly than $1/r$. (Pf: If $\chi' = -1/r$, then $\psi$ can diverge no faster than $-\log r$. But $e^{-\chi} = r$, so  $\psi^2 e^{-\chi}$ will still vanish at the horizon.)  Now consider the $\phi$ equation (\ref{phieom}). The $\chi'/2$ term will be negligible compared to the $2/r$ term so
\be\label{newphi}
\phi'' + {2\over r}\phi' -{2q^2\psi^2\over g}\phi =0
\ee

We now claim that the vanishing of $\phi$ on the horizon requires $\psi^2/g$ to diverge at least as fast as $1/r^2$. To see this, note that if $\psi^2/g \propto 1/r^\beta$ with $\beta < 2$, then one solution to (\ref{newphi}) will be singular, $\phi \propto 1/r$, and the other will be nonzero at the horizon, $\phi \propto e^{Br^{2-\beta}}$ for some constant $B$. Neither satisfies the required boundary condition. 
So $\psi^2/g$ must diverge at least as fast as $1/r^2$. But if $g \propto r^2 \psi$, then $g\psi^2 \propto r^2 \psi^4$. This can be nonzero only if $\psi \propto r^{-1/2}$. But this is more singular than $\log r$ and we have already seen that in this case the $e^{-\chi}$ term will cause its contribution to the Schr\"odinger potential to vanish.  

The net result is that in all cases, the Schr\"odinger potential must vanish at the horizon and $\Real\sigma(\w)$  remains nonzero even at zero temperature and low frequency.

 \subsection{Frequency dependence of the conductivity}
 
 If $V(z) = V_0/z^2$ near the horizon, one can determine the frequency dependence of the conductivity for small $\w$. We will see in the next section that this is the case for both classes of  extremal hairy black holes we construct. It is also the IR behavior seen in any W-shaped potential models such as those in \cite{Gubser:2008wz}, both for $AdS$ and Lifshitz IR throats, as well as string-derived models  
\cite{Gubser:2009qm,Gauntlett:2009dn}.

The derivation follows that of \cite{Gubser:2008wz} using matched asymptotic expansions and the conserved flux of (\ref{schr}). Note that it follows from (\ref{schr}) that if we define
$ \F = iA_x^*\overleftrightarrow{\partial_z}A_x,$
\be
\partial_z\F=0.
\ee
Near the boundary at infinity,
\be
\F=-i\left(A^{(0)*}A^{(1)}-A^{(1)*}A^{(0)}\right)=\omega |A^{(0)}|^2\  \Real\sigma
\ee
Since $\F$ is $z$-independent we can also calculate it near the horizon. When $V\approx V_0/z^2$, the transmitted wave solution is a Hankel function,
\be
A_x\propto \sqrt{\omega z}H^{(1)}_\nu(\omega z), \quad \nu=\sqrt{V_0+1/4}.
\ee
One can compute $\F$ for this solution and, since it is a function of $\omega z$ and is conserved, it must be $\omega$-independent.

We can then use the method of matched asymptotics, in the limit of small $\omega$, to find $A^{(0)}_x$. The result of this matching \cite{Gubser:2008wz} is that for small $\omega$

\be
A^{(0)}_x\propto \omega^{-\nu}.
\ee 
It therefore follows that for small frequencies\footnote{Small frequency is defined as the $\omega$ for which $V(z)=\omega^2$ occurs only deep in the infared where $V\approx V_0/z^2$ and possibly near the boundary if $\lambda>1$.}

\be
\Real\sigma \propto \omega^\delta, \qquad \delta=\sqrt{4V_0+1} - 1
\ee

When the zero temperature solution has emergent conformal symmetry in the infrared, one might expect that the conductivity will be a power law.
However, this argument does not require conformal invariance in the infrared, only a particular behavior of $V$. Indeed we will find that our infrared solutions which do not restore conformal invariance still exhibit this behavior of the potential.  Note that this argument determines the leading frequency dependence for small $\w$, but not the coefficient. When conformal invariance is not restored, the coefficient is typically exponentially small.

\setcounter{equation}{0} 
\section{Extremal limit of hairy black holes}

We now restrict to the case $V(\psi) = m^2 \psi^2$ and numerically construct the $T\rightarrow 0$ limit of the hairy black holes in two cases.
The behavior of the extremal hairy black holes depend on the values of $m^2$ and $q$. We consider the case $m=0$ and the case $m^2<0$ with $q^2$ larger than some (rather small) lower bound. In the next section we will comment on other values of $m,q$. In both cases, the horizon is at $r=0$.  This was strongly suggested from the numerical results at nonzero temperature: As we cool the hairy black holes, the dimensionless horizon radius $r_+/\mu \rightarrow 0$. We will present analytic formulas for the leading behavior of the solutions near $r=0$.  This leading solution depends on a free parameter which can be adjusted so that the solution asymptotically satisfies the desired boundary condition.

\subsection{$m^2 = 0$}

This corresponds to a  marginal operator, $\lambda =3$, in the $2+1$ superconductor developing a nonzero expectation value.
 We try an ansatz
\be
\phi = r^{2+\a},\quad \psi = \psi_0 - \psi_1 r^{2(1+\a)}, \quad \chi =\chi_0 - \chi_1 r^{2(1+\a)}, \quad
g=r^2(1 - g_1r^{2(1+\a)})
\ee
We have used the scaling symmetries (\ref{rescale}) and (\ref{rescalet}) to set the coefficients in $\phi$ and $g$ to one. Substituting this into the field equations  and equating the dominant terms for small $r$ (assuming $\a > -1$),
one finds:
\be
q\psi_0 = \left({\a^2 + 5\a + 6\over 2}\right)^{1/2}, \quad \chi_1 = {\a^2 + 5\a + 6\over 4(\a + 1)}e^{\chi_o}
\ee
\be
g_1 =  {\a + 2\over 4} e^{\chi_o}, \quad \psi_1 = {q e^{\chi_o} \over 2(2\a^2 + 7\a +5)}\left({\a^2 + 5\a + 6\over 2}\right)^{1/2}
\ee

\begin{figure}
\begin{center}
\includegraphics[width=.45\textwidth]{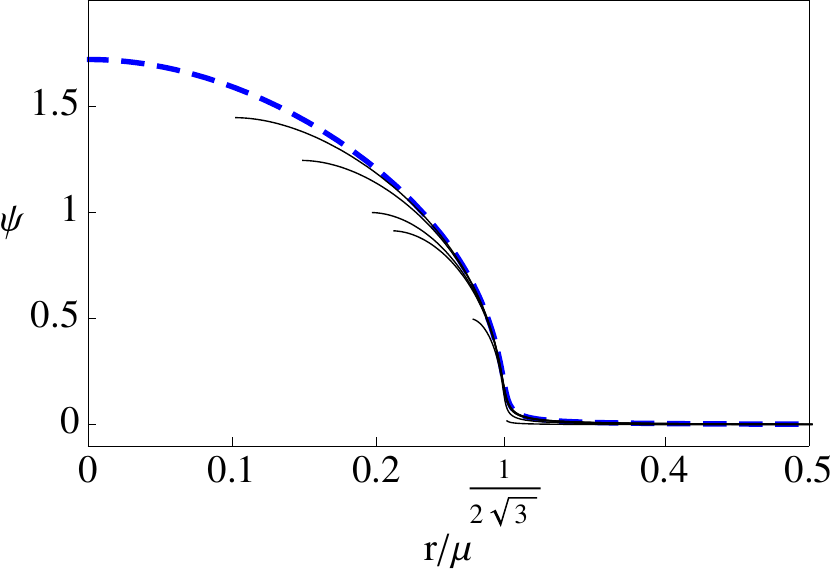}\hspace{0.2cm}
\includegraphics[width=.45\textwidth]{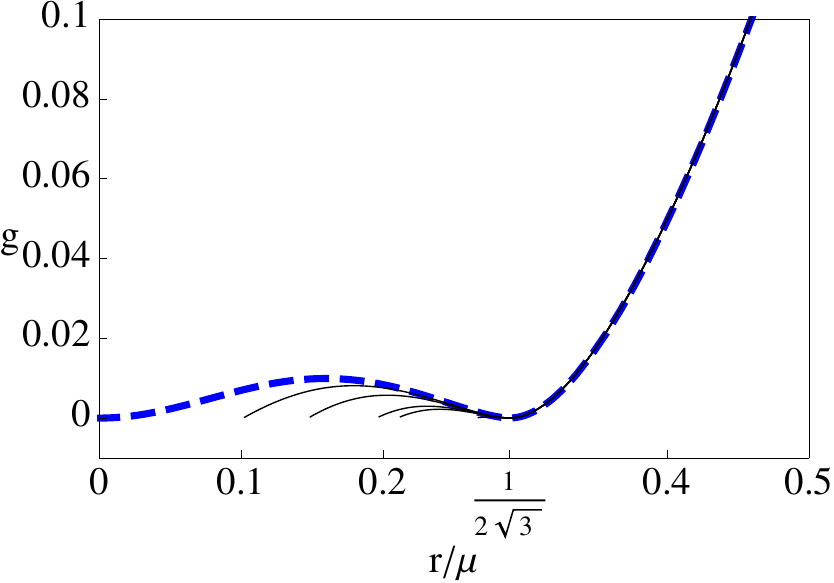}\caption{Zero temperature, $\lambda = 3$ and $q=1$ solution (dashed blue), compared to successively lower temperature hairy black holes (solid black.) Note that $g$ almost has a double zero at $r/\mu=1/2\sqrt{3}=r_+(T=0)$ for RN-AdS and all of the scalar hair is behind it. In the limit $q\rightarrow \sqrt{3}/2$ the solution becomes extremal Reissner-N\"ordstrom with all hair behind the horizon.}\label{d3cond}
\end{center}\end{figure}

 One can now numerically integrate this solution to large radius and adjust $\a$ so that the solution for $\psi $ is normalizable. One finds that this is possible provided $ q^2 > 3/4$. This is consistent with the  condition for instability (\ref{unstable}). Figure 3 shows the results for $\psi$ and $g$ for the zero temperature solution, and shows how the $T>0$ solutions approach it as $T\to 0$. The value of $\a$ depends weakly on $q$ (see Figure 4). In all cases, $| \a| < .3$. As $q\rightarrow \infty$, $\alpha=.10+.65/q^2+\cdots$ 
 
 Near $r=0$, $\chi$ approaches a constant and $g=r^2$. Thus the metric approaches $AdS_4$ with the same value of the cosmological constant as infinity. The extremal horizon is just the Poincare horizon of $AdS_4$. The scalar field approaches a constant and the Maxwell field vanishes. In terms of the dual field theory, this means that the full conformal symmetry is restored in the infrared. This is similar to the recent embeddings of holographic superconductors in string theory \cite{Gubser:2009gp}, except in those cases, the potential had more than one extrema, and the near horizon AdS radius was different from the asymptotic one.

\begin{figure}
\begin{center}
\includegraphics[width=.7\textwidth]{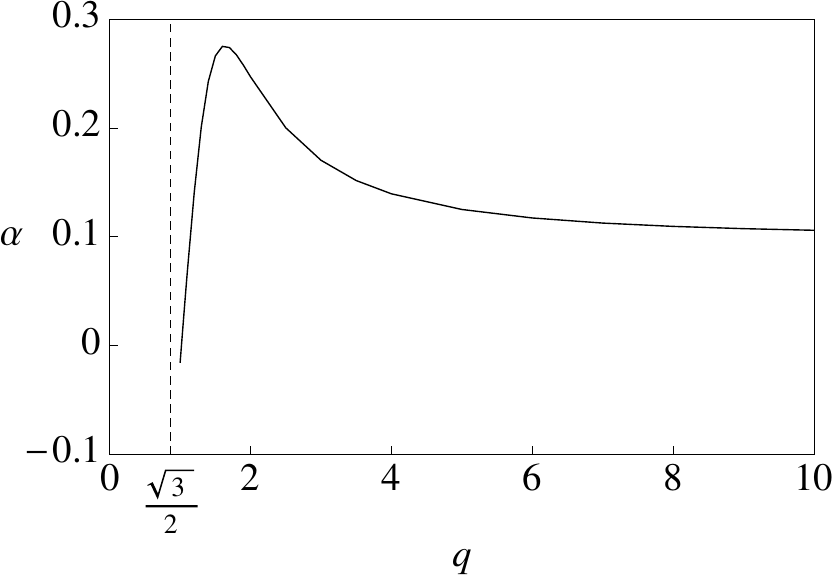}\caption{Values of $\alpha$ for various charges. Note that it approaches a constant as $q\rightarrow \infty$.}\label{pot}
\end{center}\end{figure}

These solutions are not singular at $r=0$ since all curvature invariants remain finite. 
They can be viewed as  static, charged solitons, where the electrostatic repulsion is balancing the gravitational and scalar attraction. One might wonder why the soliton does not fall through the Poincare horizon. The answer is that there is an identical copy of the soliton on the other side of the Poincare horizon which repells it. More globally, one can view the solution as describing a single soliton, localized around the Poincare horizon.

Even though these solutions are not singular, when $\a \ne 0$ they are not $C^\infty$ across $r=0$. Some derivatives of the curvature will blow up. However, there is a
 special value of the charge, $q=1.018$,  where $\a =0$. This solution is completely smooth across the  horizon.

The infrared Schr\"odinger potential in this case is easily seen to be 

\be
V_{IR}(z) =\frac{6+5\alpha+\alpha^2}{z^2},
\ee
and by the argument in section 3.3, $\Real\sigma\propto \omega^\delta$ with
\be
\delta = \sqrt{25 +20\alpha+4\alpha^2} -1
\ee
A plot of this potential for the case $q=1$ was given in Figure 2.
 
\subsection{$ q^2 > |m^2|/6 \ \ \ (m^2 < 0)$}

We try the following  ansatz near $r=0$:  

\be\label{qansatz}
 \psi=A (-\log r)^{1/2},\quad ~g=g_0 r^2 (-\log  r),\quad ~\phi=\phi_0r^\beta (-\log  r)^{1/2}
\ee
where we have used the radial scaling symmetry (\ref{rescale}) to set an arbitrary length scale in the logarithm to one. 
The behavior of $\chi$ is determined by (\ref{chieom}) and whether we expect $r\psi'^2$ or $rq^2\phi^2\psi^2e^\chi/g^2$ to dominate. 
 We assume $r\psi'^2$ dominates. (In the following section we will discuss the other choice.) Thus
\be e^\chi= K(-\log r)^{A^2/4} \label{large_q_chi}
\ee
with $K$ a constant of integration. The second term in (\ref{chieom}) is indeed negligible provided $\beta >1$.
Equating the dominant terms in the equations of motion leads  to:
\be\label{largeqconst}
 A=2, \quad g_0=-\frac{2}{3}m^2, \quad \beta=-\frac{1}{2}\pm\frac{1}{2}\left( 1-\frac{48 q^2}{m^2}\right)^{1/2}
 \ee

It is clear that for appropriate metric signature we need positive $g_0$ which tells us this ansatz is only appropriate for negative $m^2$. Since we require $\beta>1$, only the plus sign in (\ref{largeqconst}) is allowed and we require
 \be q^2>-m^2/6\ee 
  With these restrictions, our near horizon solution is
 \be\label{qansatzz}
 \psi=2(-\log r)^{1/2},\quad ~g= (2m^2/3) r^2 \log  r,\quad   e^\chi=-K\log r
\ee
\be
~\phi=\phi_0r^\beta (-\log  r)^{1/2},
\ee

\begin{figure}
\begin{center}
\includegraphics[width=.45\textwidth]{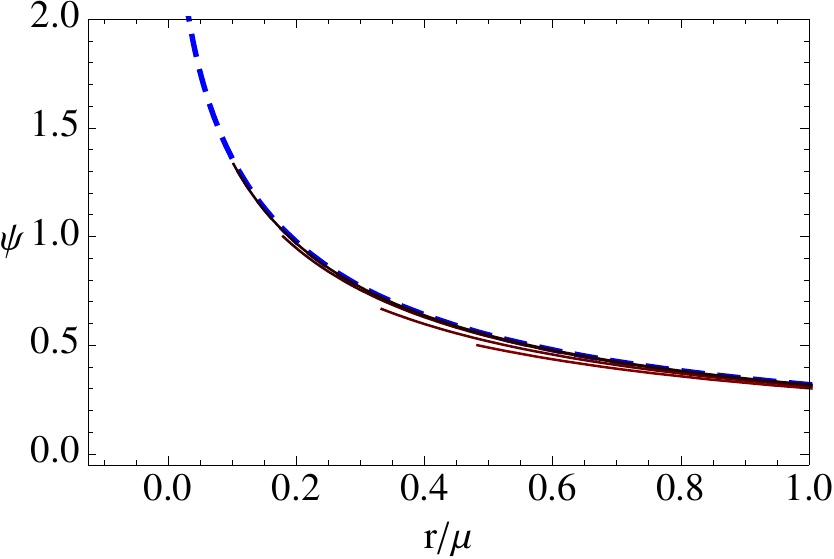}\hspace{0.2cm}
\includegraphics[width=.45\textwidth]{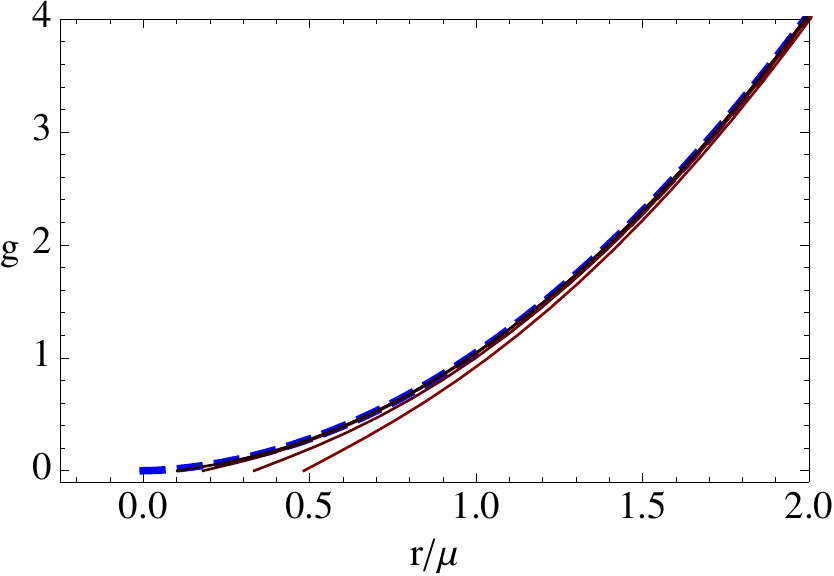}\caption{Zero temperature solution with  $m^2=-2,~\lambda=2,~q=10$ (dashed blue) compared to successively lower temperature hairy black holes (solid black.)}\label{d3cond}
\end{center}\end{figure}

The one remaining  free parameter is $\phi_0$. This parameter must be tuned so that  $\psi^{(3-\lambda)}=0$.
As an example of this ansatz, we have numerically found backgrounds for $\lambda=2$ at various values of $q$ (see Figure 5). Once again, one can see the $T>0$ solutions approach the zero temperature solution as $T\to 0$. We have also found solutions with $\lambda = 1$. One of the open questions in \cite{Hartnoll:2008kx} was whether the dimension one condensate always has a finite limit as $T\to 0$ for all $q$. Since the zero temperature solution has finite $\psi^{(1)}$ for all $q$, the answer is clearly yes.

The horizon at $r=0$ has a mild singularity. The scalar field diverges logarithmically and the metric takes the form (after rescaling $t$)
\be\label{nhmetric}
ds^2 = r^2(-dt^2  + dx_idx^i) + {dr^2\over g_0 r^2(-\log r)}
\ee
Notice that Poincare invariance is restored near the horizon, but not the full conformal invariance. (Some early indications of emergent Poincare symmetry were found in \cite{Gubser:2008pf}). From the definition of the temperature (\ref{temp}), it is clear that this solution has zero temperature.
Introducing a new radial coordinate, $\tilde r= -2(-\log r)^{1/2}/g_0^{1/2}$ the metric becomes
\be
ds^2 = e^{-g_0\tilde r^2/2}(-dt^2  + dx_idx^i) + d\tilde r^2
\ee
near the horizon which is now at $\tilde r = -\infty$. 

 Recall that  a singularity at constant radius (in a static spacetime) can be either timelike or null.  They are distinguished by looking at the behavior of radial null geodesics. If the geodesic hits the singularity in finite time, the singularity is timelike. Otherwise it is null. It is clear from (\ref{nhmetric}) that the singularity at the extremal horizon  is null.

The infrared Schr\"odinger potential in this case is easily seen to be 
$V_{IR}(z) =12q^2/|m^2|z^2$, and $\Real\sigma\propto \omega^\delta$ with
\be
\delta = \sqrt{\frac{48q^2}{|m^2|}+1 } -1.
\ee
This is always larger than two for this class of solutions. An example of the Schr\"odinger potential for $m^2 = -2, \ q=10$ was shown in Fig. 1.

\setcounter{equation}{0}
\section{Comments on other cases}

In this section we consider other choices for the mass and charge of the scalar field, still assuming a simple $m^2\psi^2$ potential. We will present  near horizon solutions, but they do not have a free parameter to adjust to satisfy our asymptotic boundary conditions. We believe that there is a subleading (possibly nonanalytic) branch of solutions which will have the missing free parameter, but we have not been able to show this. In other words, we believe the solutions below describe the universal behavior for solutions at small $r$. One piece of evidence is that in one case we can show that the low temperature behavior of the nonextremal black holes approach our solution.

\subsection{$0< q^2 < |m^2|/5 \ \ \ (m^2 < 0)$}

We now keep the same ansatz  (\ref{qansatz}) but assume that
 $rq^2\phi^2\psi^2e^\chi/g^2$ dominates (\ref{chieom}). This yields
\be
e^{-\chi}=K+\frac{q^2\phi_0^2A^2}{g_0^2}\frac{r^{2\beta-2}}{2\beta-2}
\ee
We clearly require $\beta >1$, or else the second term would be negative and diverge for small $r$. We also require the constant of integration, $K$, to vanish since otherwise the 
$r\psi'^2$ term in (\ref{chieom}) would dominate for small $r$: If $e^\chi\sim 1/K$, then
$rq^2\phi^2\psi^2e^\chi/g^2\sim{r^{2\beta-3}/K}$, while $r\psi'^2\sim 1/r\log r$.

Setting $K=0$, the leading terms in the equations of motion yield
\be
\beta=-\frac{1}{2}\left(1+{m^2\over q^2}\right)\pm\left(\frac{m^4}{4q^4}+\frac{m^2}{2q^2} -\frac{15}{4} \right)^{1/2}
\ee
\be
A^2=1-\frac{m^2}{q^2}\pm\frac{\left[ (m^2-3q^2)(m^2+5q^2)\right]^{1/2}}{q^2}
\ee
\be g_0=A^2q^2/\beta \ee

We can study either branch, and  find that requiring $\beta>1$ and $A^2>0$ yields

\be+: m^2<0, \quad 0<q^2\le-\frac{m^2}{5}
\ee
\be-: m^2<0,~-\frac{m^2}{6}<q^2\le-\frac{m^2}{5}
\ee
In the small $q$ limit, 
\be \b = -{m^2\over q^2}, \quad A^2 = -{2m^2\over q^2}, \quad g_0 = 2q^2
\ee
It may seem surprising that we started this subsection by assuming that a term proportional to $q^2$ dominated the $r\psi'^2$ term and are claiming that this is valid for arbitrarily small $q$. But this is indeed the case. 

The two branches of solutions clearly agree at $q^2 = -m^2/5$ since the terms in the square root vanish. More interesting is the fact that on the minus branch, the limit as $q^2$ approaches  $-m^2/6$ agrees with the $q^2 = -m^2/6$ solution in the previous section: $A=2$ and $\beta = 1$ in both cases. This suggests that the zero temperature solutions may form a swallowtail for $-m^2/6< q^2 <-m^2/5$. However, in this small charge case, we do not  have a free parameter to adjust to satisfy our asymptotic boundary condition on $\psi$. It may appear that we still have the freedom to adjust $\phi_0$. However, since $e^{-\chi/2}\sim\phi_0$, when we rescale our solutions so that $\chi_{new}(\infty)=0$, this shifts $\phi\rightarrow\phi e^{\chi_\infty/2}$ which cancels $\phi_0$ out of the solution, implying that all solutions are equivalent. When extended to infinity, this solution does not have $\psi^{(3-\lambda)}=0$.

It is still possible that there is a subleading branch of solutions with a free coefficient that can be adjusted to set $\psi^{(3-\lambda)}=0$. This is still under investigation. If so,  Poincare invariance would no longer be restored at small $r$ since $e^{-\chi}$ is now a power law and not logarithmic. One might expect a qualitative change in the extremal limit at very low charge since the argument for scalar hair formation is qualitatively different there: At large $q$, the instability is  dominated by the $q^2\phi^2\psi$ interaction in (\ref{psieom}), whereas at small $q$ it is dependent on the mass being below an effective BF bound in some near-horizon throat geometry.

\subsection{$q=0 \ \ \ (m^2 < -3/2)$}

The case $q=0$ is of special interest. In this case, the charge must remain on the black hole in the extremal limit. The equations simplify considerably in this case. Two of them are immediately integrated: (\ref{chieom})  implies
\be
\chi = -\int r\psi'^2
\ee
and (\ref{phieom}) implies
\be
e^{\chi/2} r^2 \phi' = \rho
\ee
where the constant $\rho$ is the charge density.
Substituting this into the remaining equations and equating the leading divergences yields
\be\label{zeroq}
\psi ={\rho\over \sqrt{-2m^2 }r^2}, \quad g= {m^4\over 2\rho^2} r^6
\ee

\begin{figure}
\begin{center}
\includegraphics[width=.6\textwidth]{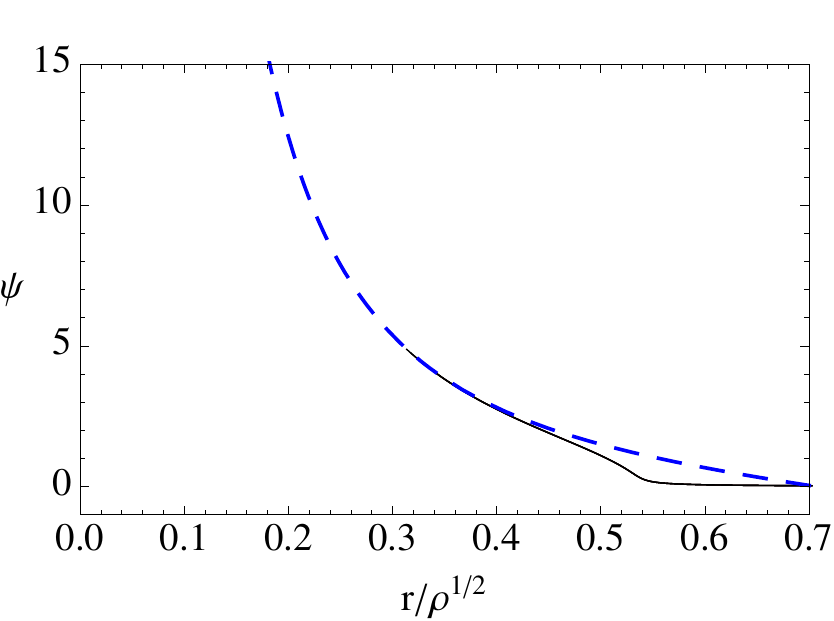}
\caption{Zero temperature zero charge ansatz, at $m^2=-2,~\lambda=2$ (dashed blue) compared to small finite temperature black hole (solid black). The radial direction is normalized with respect to $\rho^{1/2}$ rather than $\mu$ as in previous figures since the near horizon solution only determines $\rho$.}\label{d3cond}
\end{center}\end{figure}

One can show that the equations admit a power series solutions in which $\psi$ and $g$ get corrections of the form $\sum c_n r^{4n+2}$. Unfortunately, 
the leading order behavior near the singularity is uniquely determined. Once again, there is no shooting parameter. If one evolves this initial data out to large radius, one finds that it does not satisfy our boundary condition.  It turns out that there is a subleading nonanalytic component to the solution. This can be found by linearizing the equations about the leading solution given above. The result is that\footnote{Recall that $m^2 < 0$ so these perturbations vanish rapidly near $r=0$.} 
\be
\delta \psi = A e^{\rho^2/4m^2r^4}     \qquad \delta g ={|m^2|^{3/2}\over \sqrt 2\rho}r^4\delta\psi
\ee
Hopefully,  A can be adjusted to satisfy the boundary conditions at infinity. Figure 6 compares a low temperature solution for $\psi$ with the zero temperature ansatz (\ref{zeroq}). Considering that our zero temperature ansatz has no free parameters, the agreement strongly suggests that solutions will exist with our near horizon behavior.

This is the most singular of the solutions we have discussed since $\psi$ diverges like $1/r^2$. It is easy to check that the singularity is again null.

\subsection{$m^2>0$}

For $m^2 > 0$, there is a Lifshitz solution at small $r$.  This geometry was first discussed in \cite{Kachru:2008yh} in the context of a different bulk Lagrangian. (It has recently been shown that it is difficult to realize Lifshitz solutions in string compactifications \cite{Li:2009pf}.)
Consider the ansatz:
\be
\psi = \psi_0,\quad \phi= \phi_0r^z, \quad g = g_0 r^2, \quad e^\chi = r^{2-2z}
\ee
Substituting into the field equations, one finds a solution with
\be\label{lifshitz}
m^2 ={2z-2\over z} q^2
\ee
The coefficients take the form
\be
g_0 = {6\over (z+2)(z+1)}, \quad \psi_0^2 = {z\over q^2} g_0, \quad \phi_0^2 = {2(z-1)\over z} g_0
\ee
It is clear from (\ref{lifshitz}) that these solutions always have $m^2 < 2q^2$. So by scaling up $m,q$, the RN AdS solution will be unstable, and one expects that the zero temperature solution has a near horizon geometry which is this Lifshitz solution.

The leading order solution has no free parameter, so we again expect a subleading branch of solutions. This can be found by studying linearized corrections to the Lifshitz geometry which are vanishing as $r\rightarrow 0$. These linearized corrections go as

\be
\delta \psi\propto r^{\alpha_\psi},~\delta \phi\propto r^{z+\alpha_\psi},~\delta g\propto r^{2+\alpha_\psi},~\delta \chi \propto r^{\alpha_\psi},
\ee
where $\alpha_\psi$ is one of the roots of 

\be
3 z \alpha_\psi  \left[\alpha_\psi ^3+ \alpha_\psi ^2(2z+4)+ \alpha_\psi(10z-z^2) -2 z^3 + 2z^2 +8 z -8 \right]
\ee
$$+ 4 q^2 (1+z)^2 \left(z^2+z-2\right) =0.$$
It is worth noting that, for values of $q^2$ and $m^2$ satisfying (\ref{unstable}), the  perturbations always have $\Imag \alpha_\psi\neq0$ \cite{Gubser:2009cg}. This implies that the  static perturbations have oscilliatory behavior, and may imply that the Lifshitz solutions are unstable.

If solutions exist which interpolate between these Lifshitz near horizon geometries and $AdS_4$, the infrared Schr\"{o}dinger potential would be (with the new radial coordinate $x$ instead of $z$ to avoid confusion)

\be
V_{IR}(x) =\frac{2}{ x^2},
\ee
Surprisingly, this is independent of the Lifshitz  scaling parameter $z$, and has the same fall-off as the case of emergent conformal or Poincare symmetry.
From section 3.3, $\Real\sigma\propto \omega^2$.
Even more surprising is the fact that this simple result is independent of the scalar potential, so long as it admits a Lifshitz solution. This can be shown using the fact  that in our ansatz, $\psi_0$ and $\phi_0$ are determined in terms of $g_0$ by equations (\ref{phieom}) and (\ref{chieom}) which do not depend on the scalar potential.\footnote{Of course not every scalar potential admits a Lifshitz solution, and one must check that  the remaining two equations of motion allow for such a solution.}

\setcounter{equation}{0}
\section{Discussion}

We have studied holographic superconductors in the zero temperature limit. We found a simple formula for the conductivity in terms of the reflection coefficient in a one dimensional Schr\"odinger problem.  This formulation is very general. In any bulk theory in which the quadratic action for the Maxwell field takes the standard form, the conductivity will be given by (\ref{cond}).  Since the Schr\"odinger potential is bounded and vanishes at the horizon, it follows that the conductivity is nonzero at low frequency, even at zero temperature. This suggests that these models do not have an energy gap for charged excitations.  One might wonder if this is a result of the fact that our boundary theory has a global $U(1)$ symmetry which is spontaneously broken by the condensate, and hence has  a massless Nambu-Goldstone boson. A key unresolved issue is whether this survives to leading order in the large $N$ expansion which is dual to the classical gravity calculation we have done in the bulk\footnote{We thank S. Hartnoll for a discussion on this point.}.

Although we have focussed on $AdS_4$ boundary conditions, we expect a similar expression for the conductivity in terms of a reflection coefficient in higher dimensions. We also expect a similar conclusion about the absence of a hard gap. Using  $SU(2)$ gauge fields in the bulk to realize superconductivity is not likely to change this conclusion.  Even in the probe limit, $\Real\sigma(\omega)$ was not strongly suppressed at low frequency and low temperature \cite{Gubser:2008wv}, and the specific heat is a power law \cite{Peeters:2009sr} indicating the absence of an energy gap for charged excitations. To obtain a superconductor with a hard gap,  one might consider nonminimally coupled scalars in the bulk \cite{Franco:2009yz}.

We have also constructed the extremal limit of certain hairy black holes dual to the boundary superconductors.  For fixed $q$, the behavior of the zero temperature solutions changes qualitatively as one increases $m^2$ from a negative value to zero. In fact,  $m^2 =0$ might be viewed as a quantum critical point: It is only in this case that the solution has emergent conformal symmetry in the infrared. Of course, this is different from a standard quantum critical point since $m^2$ controls the dimension of the condensate in the superconductor, which is not usually viewed as a tunable parameter.

It is interesting to ask what happens to our zero temperature solutions as $m,q$ approach $ m^2 - 2q^2 = -3/2$ where the extreme Reissner Nordstrom AdS black hole becomes stable. Consider the case $m=0$. For  $q^2$ close to $ 3/4$, the metric function $g(r) \propto r^2$ for small $r$, then dips down and has a local minimum at a nonzero radius before continuing to grow as $r$ increases. As $q^2$ approaches $3/4$, the local minimum drops to zero becoming a new degenerate horizon. All the scalar hair is concentrated inside this horizon and the solution becomes the standard extremal Reissner-Nordstrom AdS metric.  We expect the same behavior for the case $m^2 < 0 $. 

The charge (per unit volume) on the black hole is given by $Q/V = e^{\chi/2} r^2 \phi'|_{r=0}$. Evaluating this in the various cases above, one finds that $Q = 0 $ for all cases except the neutral scalar, which cannot carry any charge. This can be understood as follows: Since the horizon is shrinking to zero size as $T\rightarrow 0$, any charge on the black hole would produce an infinite electric field. If the scalar has any nonzero charge, the strong electric field will pair create charged particles which will neutralize the black hole.

Suppose one perturbs  the extremal Reissner-Nordstrom AdS black hole when it is unstable. It cannot evolve to the zero temperature solutions we find since that would violate the area theorem. What happens? The point is that evolution is always at fixed energy, not fixed temperature. So the extreme Reissner-Nordstrom AdS black hole will evolve to a nonextreme hairy black hole with larger entropy. If one instead considers solutions at fixed temperature and slowly lower the temperature, one reaches the extremal black holes we have discussed.

\vskip 1cm
\centerline{\bf Acknowledgements}
\vskip .5cm
It is a pleasure to thank S. Hartnoll, H. Liu, A. Nellore, S. Shenker, and T. Wiseman for discussions. G. H. thanks the Aspen Center for Physics where some of this work was done. This work was supported in part by NSF grants PHY-0555669 and PHY-0855415.

\end{document}